# H-L PS: A Hybrid Approach for User's Location Privacy in Location-based Services

Sonia Sabir, Inayat Ali*, and Eraj Khan

***Abstract*—Applications providing location-based services (LBS) have gained much attention and importance with the notion of the internet of things (IoT). Users are utilising LBS by providing their location information to third-party service providers. However, location data is very sensitive that can reveal user's private life to adversaries. The passive and pervasive data collection in IoT upsurges serious issues of location privacy. Privacy-preserving location-based services is a hot research topic. Many anonymisation and obfuscation techniques have been proposed to overcome location privacy issue. In this paper, we have proposed a hybrid location privacy scheme (H-LPS), a hybrid scheme mainly based on obfuscation and collaboration for protecting user's location privacy while using location-based services. Obfuscation naturally degrades the quality of service but provides more privacy as compared to anonymisation. Our proposed scheme, H-LPS, provides a very high level privacy yet providing a good accuracy for most of the users. Privacy level and service accuracy of H-LPS is compared with state-of-the-art location privacy schemes and it is showed that H-LPS could be a candidate solution for preserving user's location privacy in location-based services.**

## I. Introduction

[1] The Internet of Things (IoT) is an incipient technology of Internet-based information architecture that promises a new era of technology in which every imaginable object is capable of collecting and transferring data over the Internet. IoT clasps boundless and pervasive connectivity of heterogeneous technologies, services, and standards using different devices having different platforms, architectures, capabilities, and functionalities. By combining many different technologies, services, and standards, IoT promises to provide multifolded advantages to all stakeholders in the coming years. The ascent of information technology in our everyday life has given more importance to the collection, handling and distribution of information about users. One such application area of IoT is location-based services (LBSs). LBSs have different application domains such as traffic telematics, location-based advertisements and weather forecasts, fleet management, calling taxis, routes and distance calculators, etc. All these applications allow organizations and third parties to collect and analyse data about the environment and individuals' attributes and in return offer personalized services. However, considering the increasing trends to gather more personalized and individualized data certainly raises many questions regarding individuals' privacy [2]. The high level of heterogeneity tied with the wide scale of LBSs and continuous data collection of people's private lives is considered to amplify security and privacy issues [1]. Privacy can be defined as, "Awareness and control over the collection, processing, dissemination, and use of personal data." Privacy threats are: identification, localization and tracking, profiling, Interaction and presentation, lifecycle transition, inventory attacks, and linkage [3]. IoT deployment for LBS will make data collection more pervasive, passive, and less intuitive and users will not be aware of when and where they are being watched and tracked. This pervasive and passive localization and tracking of users will make location privacy issues even more challenging. Location privacy issues are more common in LBSs. Today enormous amount of smart devices are devised with GPS that sometimes abuse our location privacy while using LBSs. There are two main approaches to solving these issues of location privacy, i.e., location anonymity and location obfuscation [4] [5] [6].

Obfuscation techniques provide a fake location of the user to a location service provider (LSP) and hide the user's exact location so that the adversary or provider cannot use his exact location for malicious purposes. LPS is an important entity in LBS system, which cannot be trusted as well and users may want to hide their personal information from LSP. Obfuscation approaches can provide good privacy at the cost of poor service accuracy.

In location anonymity, the user sends multiple dummy locations with his original location, so that LPS should not be able to know the exact location of the user. The provider replies to all the LBS queries sent by the user. The user filters queries against his original location and discards others. This anonymization approach is quite efficient as it provides full-service accuracy and good location anonymity. However, it has many limitations which are: it produces high overhead both at LSP and user device, the extent of location privacy in this approach is directly related to the number of dummy locations sent along the original location. The privacy level in an anonymization-based approach also depends upon the selection of those dummy locations and in some cases also on the security level of a third-party anonymizer if used.

In this paper, we have proposed a hybrid technique for protecting location privacy of users in LBSs. Our H-LPS scheme is a hybrid approach based on collaboration and

Sonia Sabir and Eraj Khan were with the Department of Computer Science at the COMSATS University of Science and Technology, Islamabad, Pakistan

Inayat Ali was with the Division of Supercomputing, Korea Institute of Science and Technology Information Daejeon, South Korea

*For correspondence: falcon19khan@gmail.com*

[1]

obfuscation approach. In this scheme, the users interested in LBSs collaborate with each other in an ad hoc network. Users share there location privacy requirement and their slightly obfuscated locations with each other and elect a query user (QU), which sends the location query to LSP on behalf of the group and forwards the response received from LSP to the group of users. The user with the lowest privacy requirement is elected as QU.

This H-LPS scheme can easily be adopted in wide application areas such as, by users in dense areas like shopping malls and business centers to find a desired restaurant, a bank, or any other location in that area, and by cars/taxis on a busy road to find a desired destination in a privacy-preserving way by collaborating with others cars. One more interesting application area can be providing location privacy to the spectators of mega events like summer and winter Olympics where different spectators enjoy one of the many games at nearby pavilions. H-LPS provides K-anonymity from LSP without sending multiple dummy locations or a third party anonymizer like traditional anonymity-based techniques. The wisely calculated final obfuscated location as explained in Section 3 provides anonymity to all users in the group. Results presented in this paper show that the privacy level of the proposed scheme from the malicious user is higher than the privacy level provided by the other schemes. Moreover, our scheme provides better privacy protection against malicious service providers as well.

The rest of the paper is structured as follow: In Section 2, we have provided an in-depth review and analysis of some recent privacy-preserving LBSs techniques. A novel privacy-preserving techniques has been proposed in Section 3 and its performance analysis is presented in Section 4. Finally, we have concluded this paper and presented some open issues in Section 5.

## II. Privacy-preserving LBS techniques

To address location privacy issues in LBSs, two main approaches are used, which are the obfuscation approach and the anonymization approach. Obfuscation techniques gives the fake location of the user and hide the user's exact location so that the adversary or provider cannot use his exact location for malicious purposes. In location anonymity, the user sends multiple dummy locations with original location which confuses the provider about the exact location. Researchers have proposed many location privacy techniques that are based on any one of these two main approaches. Both of these techniques have their own advantages and disadvantages. Anonymisation-based approaches provide very good privacy however these approaches have high communication overhead. Sometimes these approaches involve a third-party anonymizer that act as a single point of failure for many users and cannot be trusted as well. On the other hand, in obfuscation-based approaches, there is also a trade-off between the quality of service and privacy level. Recent work using obfuscation and anonymization techniques is given.

### *A. Obfuscation Techniques*

In [10], to address location privacy, the authors have introduced the semantic obfuscation (S-obfuscation) technique. The proposed technique creates blurred locations using ontological classification of locations based on geographical knowledge which ensures the adversary cannot identify that it is an obfuscated location. The original location was obtained through a GPS receiver (longitude, latitude), which was then converted into the ontology in the form of a hierarchy. In ontology, the classes were constructed on the basis of subdivisions of geographic area (Australia). The hierarchy has shown the relative proximity of the obfuscated location to the original location. Objects of the level 1 class were chosen as base points for objects related to the level 2 class. A base point is selected based on geographical knowledge. The performance of this technique is measured based on prediction rate (up to which extend an adversary could identify that a given location is fake) and is compared to previous Rand and dispersion techniques. The comparison shows that the prediction rate of S-obfuscation is low as compare to other classical techniques. Authors in [7] have proposed multiple obfuscation operators for the protection of user's location information. The proposed operators are enlarge (E), shift (S) and reduce (R). Users are required to mention privacy preferences in term of privacy metric value. Based on the required level of privacy one of the operator is selected to determine the obfuscated location. In this scheme, Users set either 0 or 1 as the privacy metric value. 1 means the user needs a high level of accuracy and a low level of privacy. If the privacy metric value is '0' that means the user requires an extremely high level of privacy. These operators help in identifying the user's privacy requirements and then according to the user's requirement, the actual location of the user is obfuscated. [13] proposed an obfuscation technique for those LBSs that do not require the exact location of the user rather they require distance travel by the users. These services are helpful for fitness applications that are used to pursue user's workout in term of distance-based outdoor fitness attempts, or measuring average distance travel from one point to another, an another application is 'pay-as-you-go insurance'. These applications do not require exact location so privacy in these types of applications can be achieved by hiding the exact location without suffering from accuracy problems. The proposed solution calculates the fake path travelled by a user and maintains the distance property of the path between the travelling points by using orthogonal transformation and ellipsoidal transformation.

The proposed technique preserved the location privacy of the user without exposing exact location or path. Li et al. [15] proposed a time-obfuscation-based scheme (top-privacy) to provide location privacy by sending dummy queries to confuse the provider about user's real query. In this scheme, a user sends a dummy query in free time and sends a real query when he/she require actual LBS. The proposed algorithm 'dummy query generation algorithm' consists of two modules: the dummy query selection module and the point of interest (POI) selection module. For every dummy query, a dummy location is selected by using user's real location and offset distance to make the virtual circle for limiting the user's movement. In the virtual circle, a dummy location set is calculated by comparing the location distribution of the actual user and other cells. Locations with same distribution lies in the same set. User classify all query distribution of historical queried POIs and then assign several weights to PIOs to construct the appropriate pool. Andrés et al. [8] proposed a location privacy framework that provides the desired level of privacy using the laplacian function to distort the original location of the user. In this scheme, the privacy of a user is protected with a focus on side information adversaries might have. All the obfuscation-based techniques are summarised in Table 1.

### B. *Anonymisation Techniques*

Ben et al. [9] proposed a privacy-preserving scheme called a pseudo-location updating system. In this scheme, the first $k$1 dummy locations are selected. User gets a dummy location from history and store it in its buffer, and when the two users encounter with each other, they swap some of their dummy locations to randomise the location. Patil et al. [18] proposed a mechanism called mobile crowd that hides location information from LBS providers. All the users in the region are registered to the server. Then a subset of users continuously update the buffer by getting location services from the server. A particular user get LBS from its peer instead of a third-party server. Lei et al. [14] proposed a mechanism for location privacy about k-POI in LBSs. Anonymous region is formed by generalising user's actual location and communication-cost. A mechanism is used to reduce the communication overhead while generalising user location. A new query algorithm is proposed to make all POI that are nearby the centre of anonymise zone.

Jiangyu et al. [12] proposed a cloaking-based method for privacy preservation by dividing the cloaking area into the sub-cloaking region. This sub-cloaking region may or may not contain user's exact location. Users send query with sub-cloaked areas to the provider, which gives accuracy of services by preserving user's actual location. A k-anonymity trajectory (KAT) algorithm is proposed in [16] that preserves the location and trajectory of users in a single query and continuous query. For single query, the user selects $k$1 dummy locations through sliding window-based k-anonymity algorithm. Trajectory selection mechanism is used to select $k$1 dummy trajectories for continuous queries. A summary of these techniques is provided in Table 2. Table 3 shows a comparative summary of various techniques. The performance parameters are privacy, accuracy, communication overhead, and computation cost. Most of the anonymization techniques suffer from high communication overhead whilst most of obfuscation-based techniques suffer from service accuracy problems as shown in Table 3.

## III. H-LPS: A HYBRID LOCATION PRIVACY SCHEME

In trusted third party (TTP)-based techniques, a user requests TTP to send its query to the LBS provider on its behalf. In this way, the malicious LBS provider will not know that who has actually sent this query. These techniques are very common and easy to implement but these techniques possess a single point of failure problem and also these techniques have to trust the TTP. In these techniques, if TTP is compromised, the whole user's domain will lose its privacy. Therefore, TTP-free approaches have been proposed in which the users do not have to trust on third-party and LBS providers. Rather, the user sends wisely estimated obfuscated location to the LBS provider. In this way, the LBS provider will respond to this fake location and will not know the original location of a user. However, the user will face degradation of service accuracy in such obfuscation-based approaches. Our proposed scheme uses collaboration and obfuscation in combination to propose a more sophisticated solution for protecting location privacy. The features of our proposed solution are:

- Each user can apply for LBS services with his own privacy and service accuracy requirements
- The solution is secure against malicious participants in LBS query
- Secure against malicious LBS provider
- Ensure K-anonymity for each user from the provider
- Computationally inexpensive

### A. *System Model*

We use mobile peer-to-peer system architecture.

*1) Mobile peer-to-peer architecture:* This LBS architectural model is based on collaboration between peers that are connected in an infra structureless network. In this scheme, all the users communicate with each other via multi-hop routing or directly connected peers to obfuscate their locations in cloaking areas that satisfy their privacy requirements. In this scheme, once a user finds its cloaked area, it selects its agent randomly among the other users and sends a query to the agent with its cloaked area. The agent sends the query to the LSP on user behalf and forward the response to the corresponding user after it receives a response from the LSP. The user collects responses from all its cloaked areas from all agents and computes the correct answer. The LBS model used for our proposed scheme is shown in Figure 1. The simulation set up is based on network simulator 2 (NS2) and the energy module of NS2 is used to calculate the total energy consumption of the proposed scheme. In the given model, the list of users require LBSs to first collaborate with each other's via ad-hoc network. They elect a user with low privacy requirements as QU, which sends a single query for a specific service to the provider on behalf of the group. After getting the response from the LBS provider, QU forwards the response to all participating users, (see Figure 1).

### B. *System Modules*

The proposed scheme consists of the following three steps.

*1) Selecting QU:* In this step, each user interested in LBS broadcasts '$q$' messages in its surrounding. Each message contains its *ID*, obfuscated location coordinates $(x_i, y_i)$, request for service '$r$' and privacy level requirement of $i^{th}$ user is '$p_i$'.

$$M = \{ID, (x^i, y^i), r, p^i\}$$

TABLE I: **Summary of obfuscation techniques**

| **Ref** | *Title* | **Approach** | **Description** | **Limitation** |
|---|---|---|---|---|
| Li et al., 2016 | Time obfuscation-based privacy-preserving scheme for location-based services | Obfuscation | Periodically sends dummy query with dummy locations to confuse the service provider | Extra messages cause overhead |
| Kachore et al., 2015 | Location obfuscation for location data privacy | Obfuscation | Location obfuscation for location data privacy | Application specific |
| Elkhodr et al., 2014 | A semantic obfuscation technique for IoT | Obfuscation | Fake location selected on the basis of geographical knowledge | Relies on static geographical knowledge |
| Thirunavukkarasu et al., 2014 | An obfuscation-based approach for protecting location privacy | Obfuscation | Define operator for identifying user privacy level, and calculate fake location | Use traditional obfuscation method, service accuracy is not ensured |
| Andrés et al., 2013 | Geo-indistinguishability: differential privacy for location-based systems | Obfuscation | Obtaining fake location by adding noise to actual location | Focused on privacy level |

TABLE II: **Summary of anonymization techniques**

| **Ref** | *Title* | **Approach** | **Description** | **Limitation** |
|---|---|---|---|---|
| Lei et al., 2016 | A study of location privacy protection about k interesting points queries based on agent service | Anonymity | Anonymous region is sent for LBS, provider sends all points of interest in the region | Computational and communication overhead |
| Ben et al., 2013 | Pseudo-location updating system for privacy-preserving location-based services | Anonymity | Dummy locations are selected from history buffer, swapping history with neighbor's buffer | K-locations sent to provider (overhead) |
| Liao et al., 2015 | Protecting user trajectory in location-based services | Anonymity | Sends dummy location and dummy trajectory | Computational overhead |
| Patil et al., 2015 | Hiding user privacy in location-based services through mobile collaboration: a review | Anonymity | Some users are connected to the provider, but other users get LBS from their peers | Peer may leak its interest or location data |
| Jiangyu et al., 2014 | A cloaking-based approach to protect location privacy in location-based services | Anonymity | Divide the cloaking area into sub-areas and send one sub-cloaked area to sender | Location anonymizer bottleneck, single point of failure |

TABLE III: **Comparison of proposed techniques**

| **Ref** | **Privacy** | **Accuracy** | **Communication overhead** | **Computation cost** |
|---|---|---|---|---|
| Li et al. (2016) | Medium | High | High | High |
| Kachore et al. (2015) | High | High | Low | Medium |
| Elkhodr et al. (2014) | Low | Medium | Low | Medium |
| Chow and Mokbel (2008) | Low | Low | Medium | Low |
| Lei et al. (2016) | Medium | High | High | High |
| Ben et al. (2013) | Low | High | High | Medium |
| Liao et al. (2015) | Medium | High | Medium | High |
| Patil et al. (2015) | Low | High | High | Medium |
| Jiangyu et al. (2014) | Medium | Low | High | High |

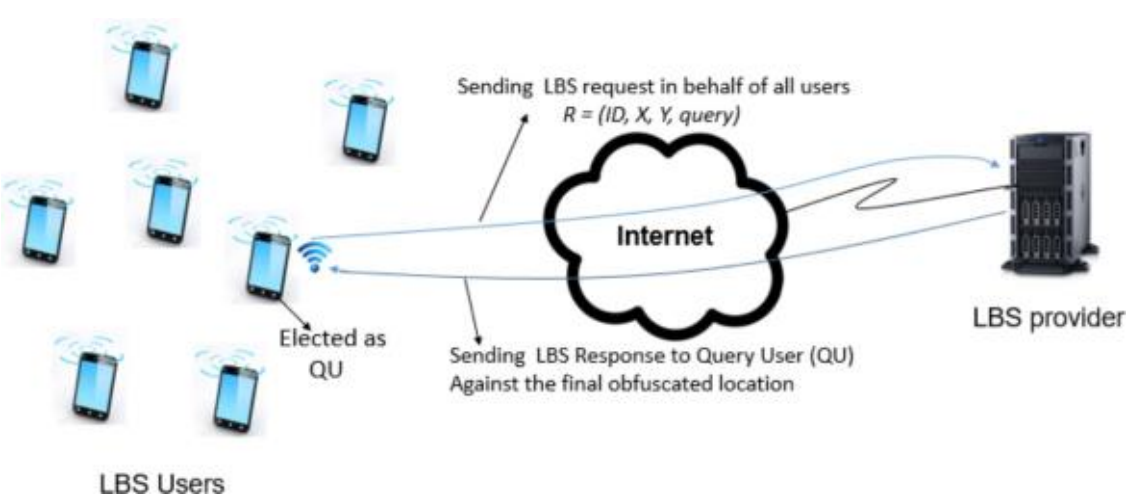

Fig. 1: LBS system

All users that have received the message and have sent a similar message will compare all the privacy levels in the received messages including their own privacy level. The user with lowest privacy requirement value p will be elected as the QU as shown by equation below and QU will then calculate efficient obfuscated location coordinates to send to the LBS provider for LBS services. The accuracy of the services will depend on the calculation of the final obfuscated location calculated by the QU and the obfuscated location send by the users. Each user will blur its location in the initial broadcast message to ensure privacy from peers. The level of obfuscation of each user in the initial query depends on the privacy level of each user. However, this obfuscation must be very small, as it is not going to affect the privacy against LSP.

$$QU = \min\{p, p^1, p^2, p^3, \ldots, p^N\}$$

where $p$ is the privacy requirement of the current user, and

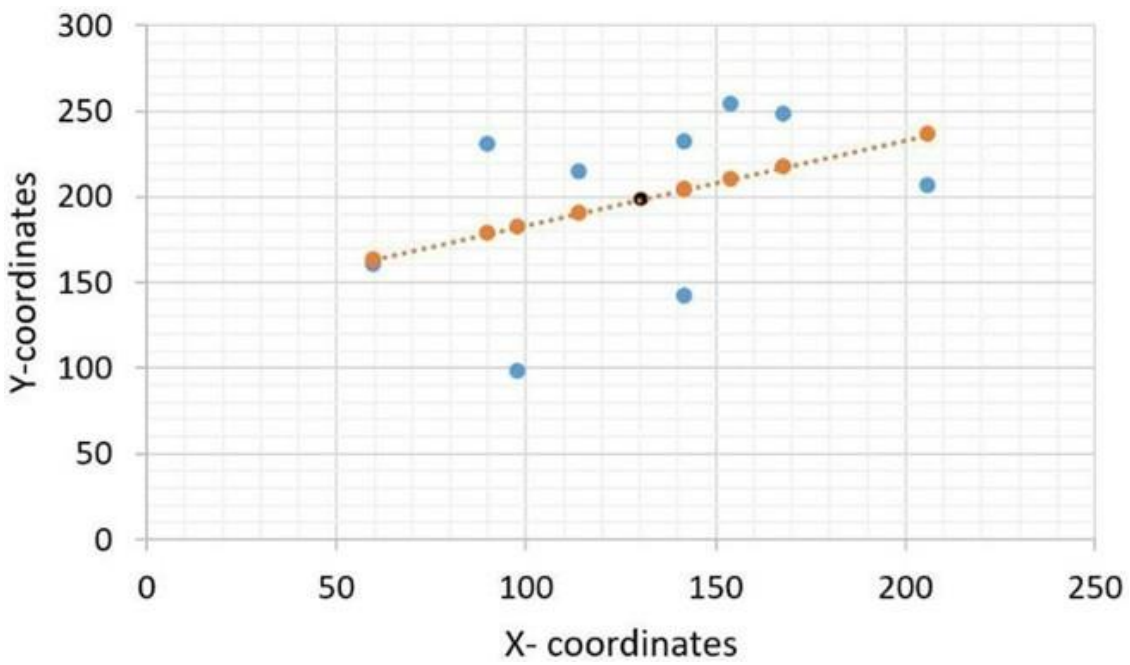

Fig. 2: Final position calculating with regression.

$p_1$, $p_2$ up to $p_N$ are the privacy requirements of user 1, 2, and $N$ respectively.

*2) Calculating Final Obfuscated Location:* After receiving the obfuscated locations of all participants, QU will first calculate the mid-point (final obfuscated location) and then will send its coordinates in the query to the LBS provider. For calculating mid-point, the QU will use regression and then will find expectation. So that the accuracy for most of the users must be satisfied. The QU will then send the coordinates determined by the mean, which is the black point on the regression line in Figure 2. The mean is taken in order to satisfy the highest service accuracy for most of the participating users as expectations will appear among the more denser area. The LBS provider will then respond to this obfuscated location, which is the location of k anonymous users. Hence, the LBS provider will not be able to identify the exact user or users group requesting for LBSs with a particular query. In this way, the proposed scheme ensures K-anonymity for each user except the QU, Moreover the QU also gets its required level of privacy with obfuscation.

*3) Forwarding the Query to the Users:* The LBS provider will process the query which represents a request from multiple users. The provider will not know exactly, whether this is a single obfuscated location of the QU or it is based on the collaboration of many users. Also knowing the ID of QU and only the midpoint of the regression line, it is not possible to find users' exact coordinates. The LBS provider will then send the response message R to the QU.

$$R = \{ID_p, ID_{QU}, response\}$$

The QU, after receiving the response, will forward the response message to all the users.

$$R_f = \{ID_{qu}, ID_i, response\}$$

where $ID_{qu}$ is the ID of QU and $ID_i$ is the ID of $i^{th}$ user.

## IV. Performance Evaluation

We have evaluated the performance of our proposed scheme against the level of service accuracy, privacy, execution time of the algorithm, energy consumption, and communication overhead. We have compared the scheme with state of the art schemes against the two main factors in any LBS scheme that is service accuracy and privacy level. We have found through simulation that our scheme provides a high level of privacy with good service accuracy for most users. Moreover, our algorithm is lightweight with low execution time and less communication overhead.

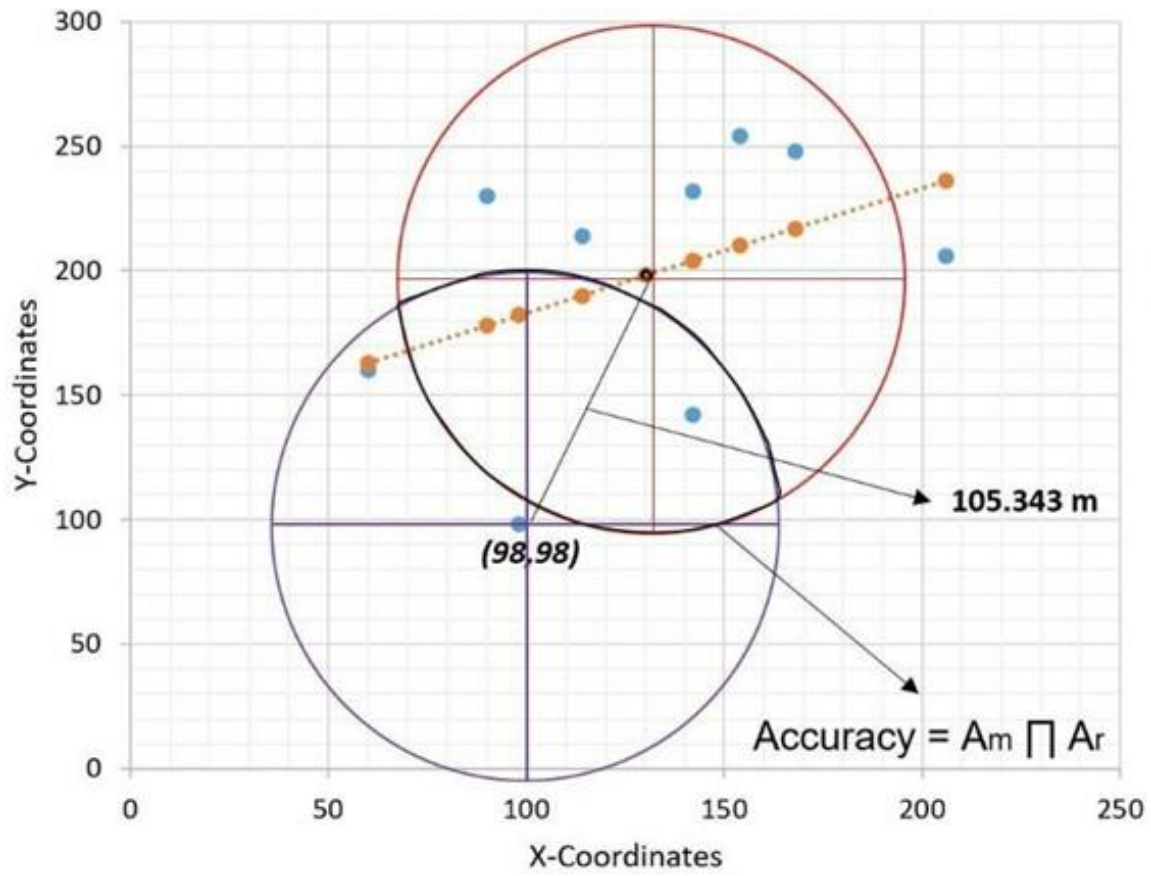

Fig. 3: Service accuracy with high privacy requirement

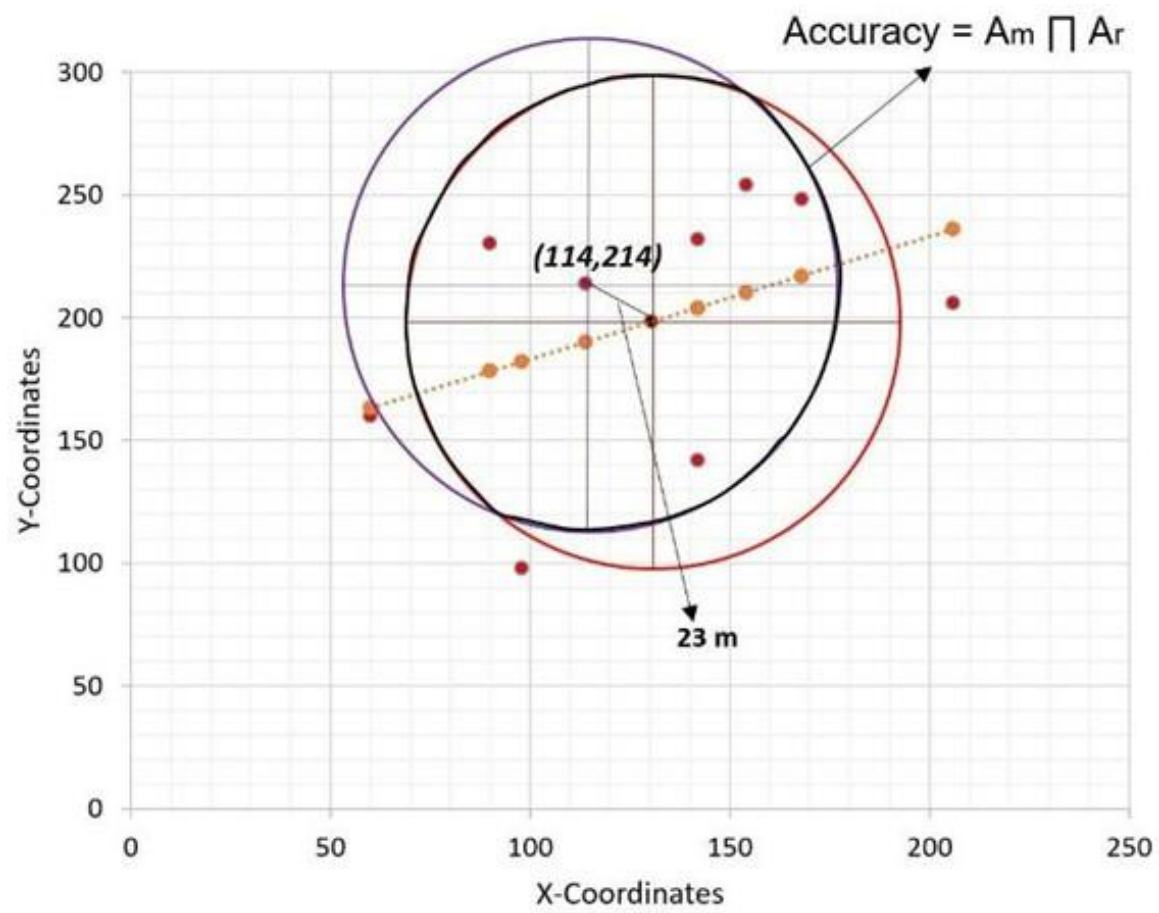

Fig. 4: Service accuracy of nearest users

### A. Service Accuracy and Privacy Level

Service accuracy and privacy level are very important parameters in LBS. However, obfuscation techniques always compromise in 100% of the accuracy. Anonymity may give 100% accuracy but it has communication overhead and involves a third party anonymiser which is not a robust system. Our solution is a blend of two approaches as discussed in Section 4, collaboration and obfuscation. The scheme provides a high level of privacy while not creating communication overhead. The scheme also provides privacy for the QU and other group users by using an obfuscated location that is obfuscated by adding random noise to its original location. The QU is sending only the mean of the regression points obtained as explained in Section 4. All the users are K-anonymous from the LBS provider. The provider will not be able to trace the motion trajectory as the group members and QU may change at every new LBS query.

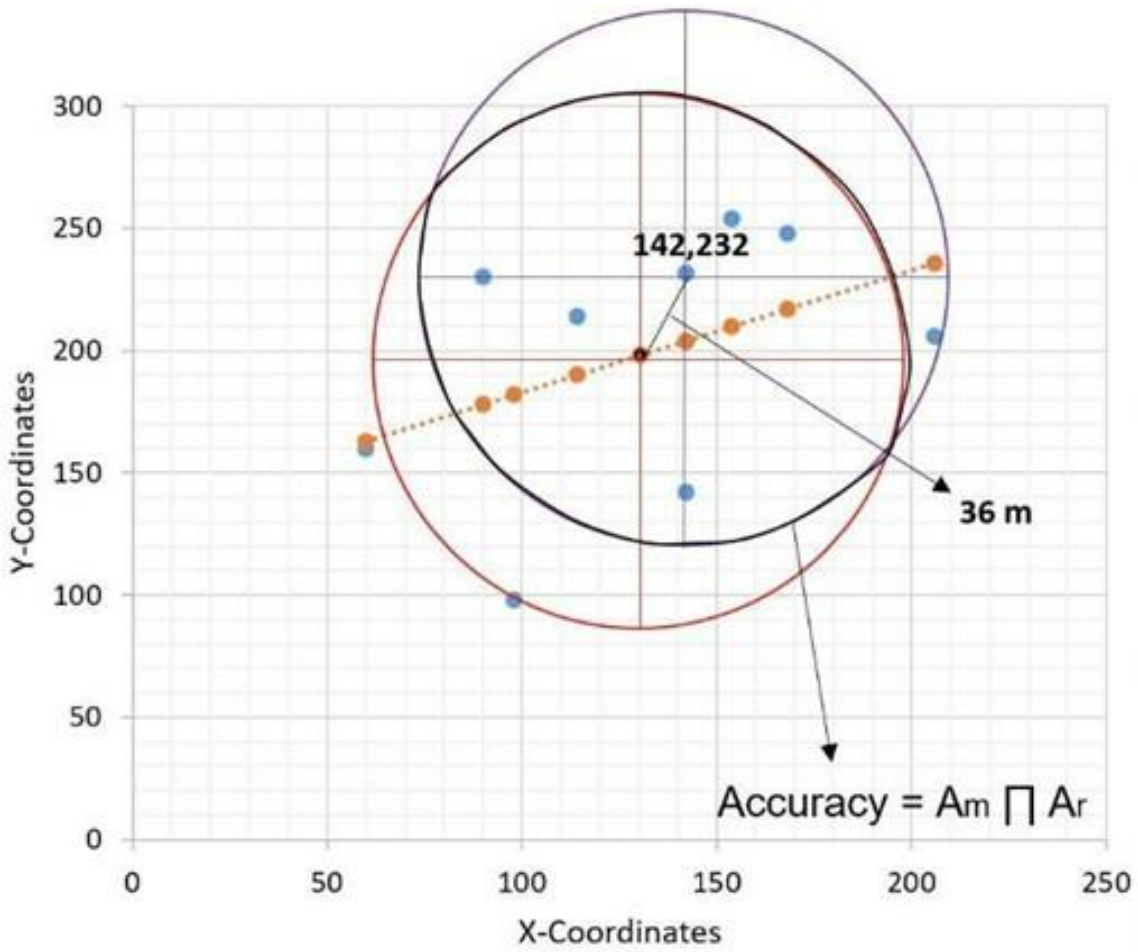

Fig. 5: Service accuracy

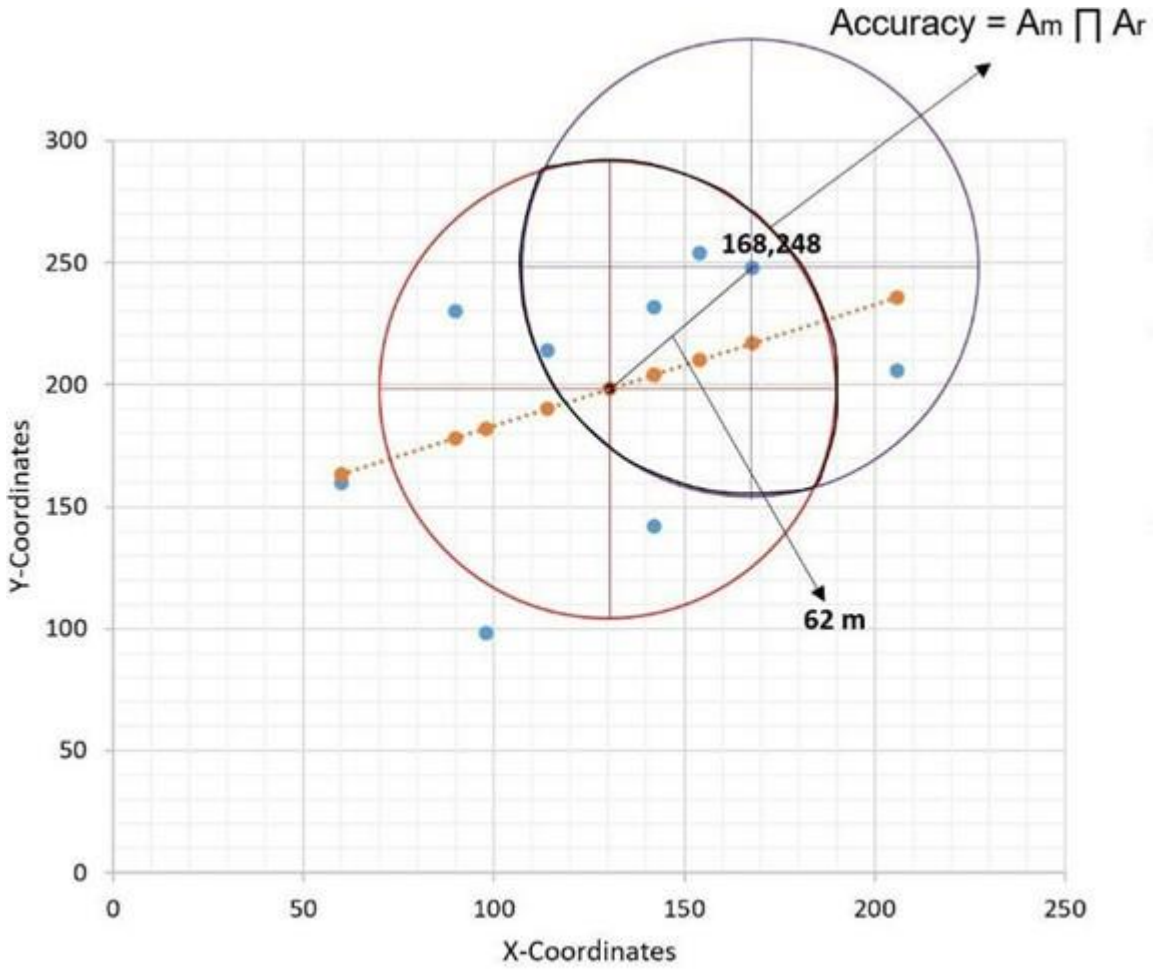

Fig. 6: Service accuracy

Our scheme allows users to have different privacy requirements by sending different privacy tag in the initial broadcast message. A user having high privacy requirements may still get the services at the cost of quality degradation as shown in Figures 3 and 6. The users that announces itself at more distance from the final obfuscated position (mean) get 25% of service accuracy with a high privacy level as shown in Figure 6. The users that are geographically near to the mean will get the highest service accuracy with excellent privacy level. Figures 3–6 show different user's service accuracy. Figure 4 shows the highest level of accuracy obtained in our simulation by the users that are near to the mean. The users end up with 80% of accuracy. Moreover, if a node appears to be at the mean or very close to the mean can get 100% of service accuracy, yet ensuring privacy from both QU and the provider.

In Figures 3–6, the overlapping areas shown by the black ellipse show the area of interest of that particular user. As different provider's applications use different ranges for providing LBSs, we in our calculations have assumed that the area served by a single query is in the circular range of 250 meters in diameter. The calculated percentage of accuracy is not dependent on the serving area range.

### B. Privacy Level

The proposed scheme provides a high level of privacy. H-LPS provide privacy from QU and LBS providers as well. Privacy level can be calculated by using entropy. The entropy H is given by the equation below.

$$H = -\sum_{i=1}^{k} p_i \log_2 p_i$$

The maximum entropy will be $\log_2 K$, where $p_i = \frac{1}{k}$ and $i = 1, 2, 3 \ldots, k$. We have calculated the entropy of our scheme to determine the privacy level of all users from the QU and among each other. We have compared the privacy level with two schemes DLS (Niu et al., 2014) and KAT (Liao et al., 2015). The results show that our scheme is performing better than the others in term of privacy yet keeping a good service accuracy. The comparisons in Table 4 shows the privacy users in H-LPS scheme from QU and from other users. KAT creates high overhead to protect users' trajectory whilst H-LPS is naturally secure from finding users' trajectories by the provider. The above results show that H-LPS have similar or higher privacy levels in comparison with other schemes. The privacy level of users among themselves and from the QUs is more than the privacy level in KAT and DLS. Moreover, the H-LPS scheme is highly privacy preserving in term of privacy preservation from the provider. The entropy in case of privacy from the provider is not dependent on the privacy level and it provides similar privacy to all users based on the highest privacy requirement of one of the participating users.

TABLE IV: **Privacy level comparison with other schemes**

| **Entropy** | *Privacy* | | |
|---|---|---|---|
| | *degree = 3* | *degree = 7* | *degree = 10* |
| DLS | 1.58435 | 2.80409 | 3.31354 |
| KAT | 1.58435 | 2.80033 | 3.30315 |
| H-LPS | 1.58496<br>≥3.3219 | 2.80778<br>≥3.3219 | 3.32193<br>≥3.3219 |

### C. Execution Time and Power Consumption

The core of the algorithm is calculating the mean of the coordinates of the location received from the group of users. This algorithm always run at the QU to calculate the final obfuscated locations that will satisfy the user's interest at the highest possible level. We have checked the execution time of our algorithm by running it multiple times and taking the average time. We have also checked the time against varying numbers of users in the group. Figure 7 shows that the execution time of our algorithm is very low in the range of milliseconds. The growth in execution time with an increasing number of users in the group is very small. The time complexity of our proposed algorithms is O(n). Figure 8 shows the very low power consumption of our scheme.

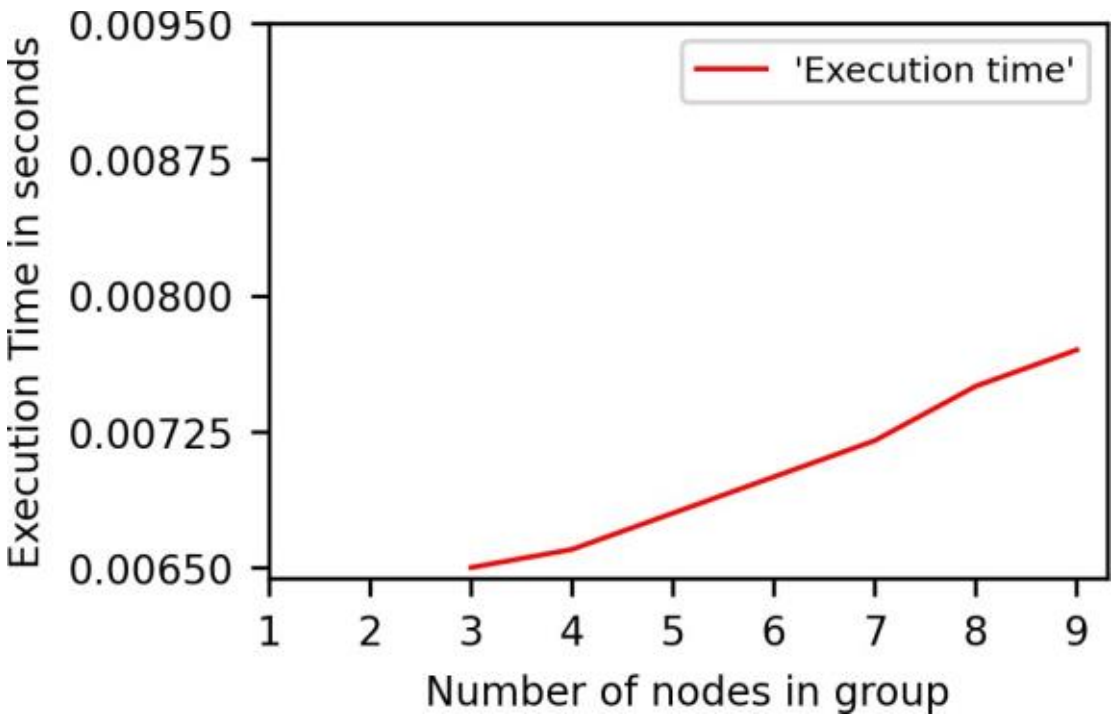

Fig. 7: Execution time of regression function

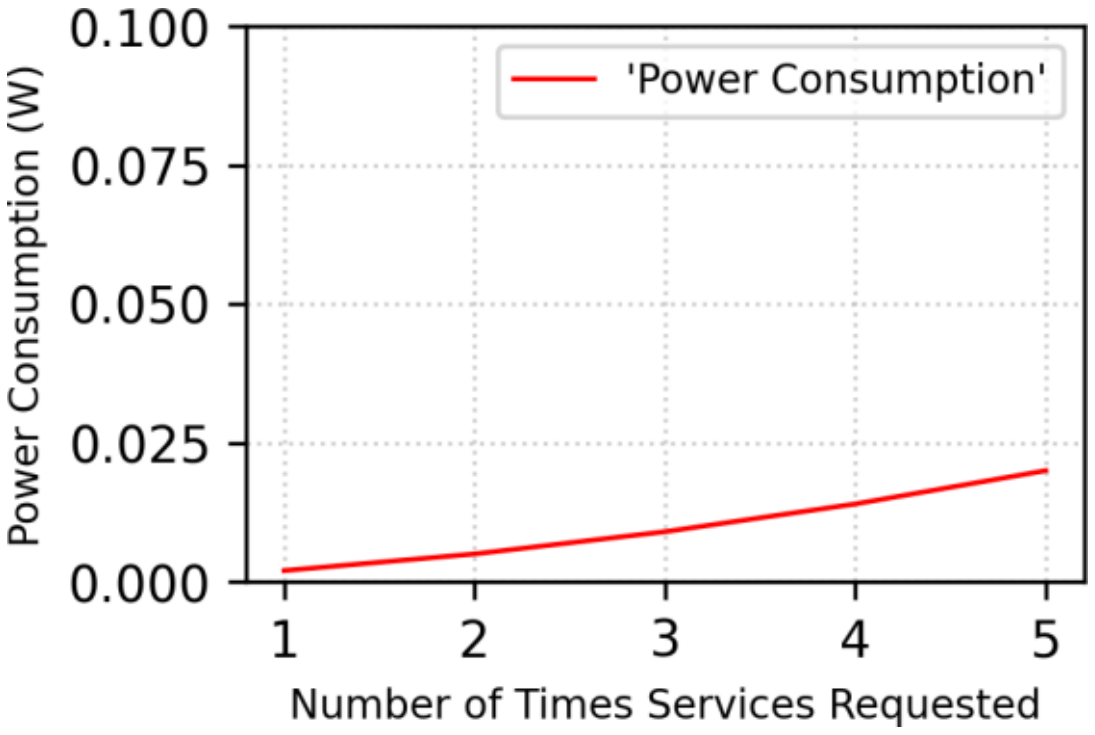

Fig. 8: Power consumption of H-LPS

## V. Conclusion

LBSs are getting more attention owing to its wide use in diverse application areas. Privacy problem is a hurdle in LBS services and need to be resolved. Many techniques have been proposed to solve the location privacy issues in LBS. Two well-known approaches, anonymity, and obfuscation have been used by different researchers to cope with these issues. However, both of the mentioned approaches have their inherited shortcomings. In this paper, we have proposed the H-LPS scheme to preserve user's location privacy in LBSs. It is a hybrid approach that merges both obfuscation and anonymity. Our proposed scheme provides a high level of accuracy from both QU and the LBS provider. H-LPS privacy level is compared with DLS, and KAT and the results show that H-LPS is quite better in term of privacy level and overhead. With our simulation results, it is shown that the proposed scheme could be a promising candidate for the state of practice in location privacy solutions. However, some open issues in this scheme to be addressed in future are: the election of QU and the trade-off between service accuracy and privacy level. QU election may also consider other parameters like the remaining energy of the QU device to improve the overall energy consumption of the scheme. Moreover, the trade-off between service accuracy and privacy level can be reduced by designing more sophisticated methods for calculating the final obfuscated location that is included in the query to the LBS provider.